\documentclass[pra,twocolumn,superscriptaddress]{revtex4}%
\usepackage{amssymb}
\usepackage{amsmath}
\usepackage{graphicx}
\usepackage{dcolumn}
\usepackage{xcolor}
\usepackage{bm}
\usepackage{subfigure}
\usepackage{amsfonts}
\usepackage{appendix}
\usepackage{tikz}
\usetikzlibrary{quantikz,backgrounds,fit,decorations.pathreplacing}
\usepackage{xcolor}%
\providecommand{\U}[1]{\protect\rule{.1in}{.1in}}

\begin{document}
\preprint{APS/123-QED}
\title{ Classification of data with a qudit, a geometric approach }
\author{A. Mandilara}
\affiliation{Department of Informatics and Telecommunications, National and Kapodistrian
University of Athens, Panepistimiopolis, Ilisia, 15784, Greece}
\affiliation{Eulambia Advanced Technologies, Agiou Ioannou 24, Building Complex C, Ag. Paraskevi, 15342, Greece}
\author{B. Dellen}
\affiliation{Faculty of Mathematics and Technology, University of Applied Sciences, Koblenz, Germany}
\author{U. Jaekel}
\affiliation{Faculty of Mathematics and Technology, University of Applied Sciences, Koblenz, Germany}
\author{T. Valtinos}
\affiliation{Department of Informatics and Telecommunications, National and Kapodistrian
University of Athens, Panepistimiopolis, Ilisia, 15784, Greece}
\author{D. Syvridis}
\affiliation{Department of Informatics and Telecommunications, National and Kapodistrian
University of Athens, Panepistimiopolis, Ilisia, 15784, Greece}
\affiliation{Eulambia Advanced Technologies, Agiou Ioannou 24, Building Complex C, Ag. Paraskevi, 15342, Greece}

\begin{abstract}
We propose a model for data classification using isolated quantum $d$-level systems or else qudits.
The procedure consists of an encoding phase where classical data are mapped on the surface of the qudit's Bloch hyper-sphere via rotation encoding,
followed by a rotation of the sphere and a projective measurement. The rotation is adjustable in order to control the operator to be measured, while 
 additional weights are introduced in the encoding phase adjusting the mapping on the Bloch's hyper-surface. During the training phase, a cost function based on the average expectation value of the observable is minimized using gradient descent thereby adjusting the weights.
Using examples and performing a numerical estimation of lossless memory dimension, we demonstrate that this geometrically inspired qudit model for classification is able to solve nonlinear classification problems using a small number of parameters only and without requiring entangling operations. 
\end{abstract}
\maketitle

With low-depth quantum circuits coming to pass, the interest for devising applications for these physical units has much increased. One fast-developing direction that already forms a sub-discipline of Quantum Machine Learning \cite{Briegel, Biamonte} is devising  methods  for addressing 
 problems of classical  machine leaning (ML) with variational quantum circuits (VQCs) \cite{VQC}. These types of quantum circuits have adjustable angles in gates which can be trained in a  fashion analogous to neural networks \cite{Svore, Farhi, Latore, Harrow}.
 On formal level though the mathematical analogy of VQCs with neural networks is far from  straightforward, mainly due to the reversibility of VQCs,
 and the problem of quantum neuron is usually approached with more intrigued
 models \cite{quest, artneu, genQNN, quaneu, quaTrann, uniqnn}.
 In addition to neural networks, VQCs  show similarities with classical kernel machines \cite{Maria1, Maria2, Harrow} by generating a feature map of classical data in the Hilbert space.
In general, the interpretation and most profitable use  of VQCs in ML tasks remains an open topic of discussion, including the accurate evaluation of their capacity and their potential or advantages compared to classical models.

This work aims to contribute to the question whether quantum circuits are suitable for solving ML tasks and how increasing the dimension of the Hilbert space can be exploited for this purpose. There are two paths to follow: One is to employ $n$ entangled qubits
  achieving  an exponential increase of space, the other, less investigated path, is to 
  employ qudits. For a single qudit, the dimension of the Hilbert space is increasing linearly with
  $d$ and without requiring entangling operations which remain demanding on  practical level. Our quantum toy model consists of a single qudit operated by  a low-depth quantum circuit (which we call single layer). With these limited resources, we are able to show that with a proper encoding  and adjustment of $d$ with respect to the dimension of the input one may achieve double lossless memory (LM) dimension \cite{LM}  as compared to habitual single-layer neural networks (NN) possessing the same number of trainable parameters. This effect cannot be achieved in the absence of parameters in the encoding phase controlling the feature map on the Bloch hyper-sphere.

  Going one step further, while keeping the input dimension fixed, 
  the capacity of the quantum system (in the sense of LM dimension) can be further increased by either re-uploading the data \cite{Latore, Heid}, this way introducing more depth into the quantum circuit, or, alternatively, as we propose in this work, to use higher-dimensional quantum systems by increasing $d$. We get preliminary evidence that the two methods give comparable results and therefore the selection should be done in dependence of the available resources.

The structure of the manuscript is as follows. We start by  introducing the Bloch hyper-sphere representation of a qudit, and then, based on this,  we develop a general scheme for mapping the data on its surface and rotating them. We then evaluate different encoding-rotation models according to  LM dimension and we draw conclusions on optimal methodology.
We  illustrate the efficiency of qubit and qutrit  models by applying them to
standard classification problems including both synthetic and real-world data.

\section{The Bloch hyper-sphere of a qudit}
A qudit stands for the state of a $d$-level quantum system just as a qubit describes a quantum $2$-level system. A qudit state `lives' in the $d$-dimensional Hilbert space which is spanned by the eigenstates of the Hamiltonian of the system. Let us denote by $\left\{\left|k\right\rangle\right\}_{k=0}^{d-1}$ such a set of normalized eigenstates. Then one can
express a generic qudit state 
\begin{equation}
\left|\psi\right\rangle=\sum_{k=0}^{d-1} c_k \left|k\right\rangle
\end{equation}
by $d$ complex amplitudes $c$ over this basis, being constrained by the normalization condition $ \sum_{k=0}^{d-1} \left|c_k\right|^2 =1$.

We claim a full  $su(d)$ algebra for  the system, spanned by $d^2-1$ generators $\left\{\hat{g}_i\right\}$ that can be chosen to be orthogonal with respect to the Hilbert-Schmidt product such that 
and $Tr\left(\hat{g}_i^\dagger\hat{g}_j\right)= G \delta_{i,j}$ with $G$ a positive constant. For $d=2$ these generators can be identified with the Pauli operators ($G=2$) while  for $k=3$ and $G=2$ with the Gell-Mann operators 
(see Appendix). Extending the set $\left\{\hat{g}_i\right\}$ by an element  $g_0= \sqrt{\frac{G}{d}} \hat{1}$,   the generators of the algebra form a basis in Hilbert-Schmidt space of Hermitian operators so that any observable $\hat{H}$ of the qudit can be written as 
\begin{equation}
\hat{H}=\sum_{m=0}^{d^2-1} h_m \hat{g}_m= h_0 \hat{g}_0+\phi \sum_{m=1}^{d^2-1} n_m \hat{g}_m=h_0 \hat{g}_0+ \phi ~\vec{n}.\vec{\hat{g}}
\label{H}
\end{equation}
 with $h_m = Tr\left(\hat{g}_m^\dagger \hat{H}\right)/G$ $\in\Re$, $\vec{n}= \left\{n_1, n_2, \ldots, n_{d^2}\right\}$ a normalized real vector and $\phi$ an angle.

The density operator $\hat{\rho}$  of a pure state $\left|\psi\right\rangle$, $\hat{\rho}=\left|\psi\right\rangle\left\langle \psi \right|$,
being a positively defined Hermitian matrix with $Tr(\hat{\rho})=1$, can also be decomposed on the basis of the generators as
\begin{equation}
\hat{\rho}=\frac{1}{G}\hat{g}_0 + \sum_{m=1}^{d^2-1} r_m \hat{g}_m
\end{equation}
with $r_m = Tr\left(\hat{g}_m^\dagger\hat{\rho}\right)/G$ and  $\vec{r}= \left\{r_1, r_2, \ldots, r_{d^2-1}\right\}$ proportional by $1/G$ factor to the unit-length Bloch vector, living on the $(d^2-2)$-dimensional surface of the so-called Bloch hyper-sphere.
 For completeness we note here that pure states occupy only a sub-manifold of this surface of dimension $d^2-d$ while the rest of the surface corresponds to non-positive Hermitian matrices. Mixed states correspond to vectors inside the Bloch hyper-sphere.

Furthermore, since any unitary operation $\hat{U}$ is generated by a Hermitian matrix $\hat{H}$ as $\hat{U}=e^{i \hat{H}}$, in view of the decomposition (\ref{H}), one can re-write $\hat{U}=e^{i \phi ~\vec{n}.\vec{\hat{g}}}$ up to a phase factor. 
The latter expression leads (with some extra work) to the interpretation  of a unitary operation acting on a pure state $ \hat{U}\left|\psi\right\rangle$ (or $ \hat{U}\hat{\rho}\hat{U}^{\dagger}$) as a rotation of Bloch vector around the $\vec{n}$-axis for an angle proportional to $\phi$.  One can also see that the most general unitary operation $U_{\vec{n}}\left(\phi\right)$ is parameterized by $d^2-1$ real parameters.

Measurable quantities on a qudit are described by Hermitian operators which, again in view of the decomposition presented in Eq.(\ref{H}), define a direction on the Bloch hyper-sphere. In addition, the $d$ eigenvectors of the observables, corresponding to $d$ real different measurement outcomes, i.e. eigenvalues, are mutually orthogonal to each other and offer a  separation of the Bloch hyper-surface into $d$ adjacent segments of equal area, in  absence of degeneracies.

\section{Employing qudits for supervised classification tasks \label{method}}
Let us consider classical data consisting of $n$ $k$-dimensional feature vectors $\left\{\vec{x}\right\}$, i.e., $\vec{x}=\left\{x_1, x_2,\ldots x_k\right\}$. Every data point belongs to one od  $M$ classes. A random subset of the data composed  of $l$-elements ($l<n$), $\left\{\vec{x}\right\}_l$,  is picked as the training set.

\subsection{Quantum resources }
For this problem, the required resource is a single qudit where $d^2-1\geq k$ and  $d^2-1- k$  increasing with the complexity of the task.
One should be able to perform the full $SU(d)$ group of operations on the qudit and in addition to measure a single observable $\hat{O}$. For simplicity,
we assume the spectrum to be non-degenerate, yielding $d$ distinct measurement outcomes. Since the classification is based on mean values of measurement outcomes, one should be able to perform experiments in identical conditions multiple times.

\subsection{ Encoding classical data }
Let us now introduce $K=S+W$ adjustable weights that we separate into two groups:
 $\vec{s}=\left\{s_1,\ldots s_{S}\right\}$ and
$\vec{w}=\left\{w_1,\ldots w_{W}\right\}$, 
with $S, W \leq d^2-1$.

In the first part, there is the encoding phase where the  classical data, i.e., the elements of the vector $\vec{x}_i$, together 
with the adjustable weights $\vec{s}$,  are ``uploaded''  on the qudit that is initially in its ground state:
\begin{equation}
\left|\psi_{\vec{x}, \vec{s} }\right\rangle= \exp\left[i x_1\sum_{j\in A_1} s_j \hat{g}_j+
 \ldots  i x_k\sum_{j\in A_k} s_j \hat{g}_j \right]\left|0\right\rangle~~. \label{enco}
\end{equation}
where $A_j$ implies different grouping of the generators with $A_j\cap A_k=0$ being a suggestive condition. With $\left|0\right\rangle$ we denote the 
ground state of the qudit.
Overall, the angles and axis of rotation of the initial vector $\left|0\right\rangle$ are related to both classical data and adjustable weights $\vec{s}$ in an intrigued way, and the result of such encoding is a map from the Cartesian space, where the inputs are initially described ($k$-dimensional real vector $\vec{x}$), 
onto the surface of the $d^2-1$ dimensional hyper-Bloch sphere. Given the requirement $k\leq d^2-1$, we actually map the data onto a higher dimensional feature space characterized by the kernel
\begin{equation}
 \left|\left\langle \psi_{\vec{x}_2, \vec{s} } \right.\left|\psi_{\vec{x}_1, \vec{s} }\right\rangle\right|^2~~.
\end{equation}
In the Appendix, we provide an explicit expression of the simplest kernel  employed in this work,  namely the qubit model A (see Table~\ref{Tab1}). 
Contrary to the  usual rotation encoding
consisting of successive rotations around orthogonal directions of the Bloch sphere, resulting in cosine kernels, the `combined' encoding
of Eq.~(\ref{enco}) results in more intrigued kernels. Naturally, the complexity of these kernels is increasing with $d$ and $k$.

\subsection{ Rotating and measuring}
After mapping the data onto the hyper-sphere, it is separated into $M$ groups.
A projective measurement of $\hat{O}$ observable, provides, with some probability which depends on the state  Eq.(\ref{enco}), an outcome from the $d$ values of its   spectrum $\left\{o_1,\ldots,o_d\right\}$ (arranged in increasing order). 

We take the habitual assumption that the whole procedure can be repeated many times in an identical way and
 use the mean value of $\left\langle \hat{O}\right\rangle$ that lies in the interval $\left[o_1,o_d\right]$ to divide the interval (equally or unequally) into $M$ segments, classifying the data, i.e.,  $\left[o_1,y_1\right],~\left[y_1,y_2\right]\ldots\left[y_{M-1},o_d\right]$. 
To get optimum results though  one should be able to rotate $\hat{O}$ in order to  `match' its orientation  with the one of the mapped data on the hyper-surface. Alternatively, one can keep $\hat{O}$ intact and    rotate $\left|\psi_{\vec{x}_l, \vec{s} }\right\rangle$.
So, in this stage, one applies arbitrary rotations to the state vector carrying the classical information, yielding
\begin{equation}
\left|\psi_{\vec{x}_l, \vec{s},\vec{w} }\right\rangle= \exp\left[i \sum_{j= 1}^{d^2-1} w_j \hat{g}_j\right]\left|\psi_{\vec{x}_l, \vec{s} }\right\rangle ,\label{rot}
\end{equation}
and measures $\hat{O}$. Let us note that it is not always profitable in terms of capacity to keep all  the weights $w_j$ in Eq.~\ref{rot} and some should be ignored or set zero so that $W\approx S$.

The whole  `encode-rotate-measure' scheme is repeated many times in identical conditions until mean value for the measurement 
\begin{equation}
\left\langle \hat{O} \right\rangle_{\vec{x}_l, \vec{s},\vec{w}}=\left\langle \psi_{\vec{x}_l, \vec{s},\vec{w} } \right|\hat{O}\left|\psi_{\vec{x}_l, \vec{s},\vec{w} }\right\rangle \label{mean}
\end{equation}
is obtained that classifies the data point $\vec{x}_l$ according to the choice of segmentation $\left\{y_i\right\}$ of the interval $\left[o_1,o_d\right]$ of mean values. The values $y_i$ can 
be also adjustable in the same way the threshold values of perceptrons in neural networks can be variable and optimizable.

\begin{widetext}
One may summarize the total scheme in the following diagram:
\begin{center}
\begin{quantikz}
\lstick{\ket{0}} & \gate{Encoding} & \phase{|\psi_{\vec{x}, \vec{s} }\rangle} & \gate{Rotation} & \phase{|\psi_{\vec{x}, \vec{s} , \vec{w} }\rangle} & \meter{} \arrow[r] & \rstick{Measurement of $\hat{O} $}\qw
\end{quantikz}
\end{center}
\end{widetext}

Finally, while the full scheme could be written as 
\begin{equation}
\left|\psi_{\vec{x}, \vec{s},\vec{w} }\right\rangle= \exp\left[i \hat{H}\right]\left|0\right\rangle ,
\end{equation}
it is important to note that $\hat{H}$ is highly non-linear in the input $\vec{x}$ due to BCH formula. Our scheme, in contrast, relies on a simple linear encoding of Eq.(\ref{enco}).

\subsection{Training}
To perform the training, we define a loss function that
penalizes misclassified data of the training set
\begin{equation}
E=\sum_{i \in T}(\left\langle \hat{O}\right\rangle_i-Y_i)^2~~,\label{error}
\end{equation}
while correctly classified data do not contribute to its value.
Here, $T$ is the set of misclassified data of the training set, and $Y_i$ is the upper or lower value of the spectral segment that characterizes  correct class for the $i$th point. In Section~\ref{rwd} we  use for  convenience the cross entropy loss function.

The optimization of parameters implies a minimization of $E$, which is achieved (in all analysis apart the examples in Section~\ref{rwd}) by gradient descent.
The landscape of $E$ though contains a number of local minima, and, when starting from a random initial point in the space of parameters $\vec{s}\vee \vec{w}$, the procedure might get trapped in one of those. To improve minimization,  we use a sample of $~l=50$  initial points and we pick the best result among all runs of gradient descent. 
When dealing with real-world-data using a qutrit (in Section~\ref{rwd}) and comparing its outcome the one obtained with classical models, a more advanced stochastic gradient descent is applied.

\section{Lossless memory dimension of different encoding-rotation models \label{LM}}
In this section, we compare different
models of encoding using a measure of capacity with clear theoretical meaning that is also suitable for numerical evaluation.   Our aim is not to accurately compare with the capacity of classical neural networks \cite{QCapacity}, 
but to identify optimum way for introducing the trainable parameters in encoding and rotating stages, Eqs.\ref{enco} and \ref{rot}, of the  proposed scheme. Due to limited computational capacity our numerical tests are not `exhaustive' but indicative.

We are employing a recently suggested measure \cite{LM}, which has been constructed for evaluating the informational/memory capacity of multi-layered classical neural networks, the so called LM dimension. This  is a generalization of the Vapnik–Chervonenkis (VC) dimension \cite{VC} that is based on the work of MacKey \cite{Mckey}, embedding the memory capacity  into the Shannon communication model. The definition of LM dimension \cite{LM} is the following:
\begin{itemize}
\item \textit{The LM dimension $D_{LM}$ is the maximum integer
number $D_{LM}$ such that for any dataset with cardinality
$n \leq D_{LM} $ and points in random position, all possible labelings
of this data set can be represented with a function
in the hypothesis space.}
\item  A set of points $\left\{ x_n  \right\}$
in $K$-dimensional space is in random position, if and only
if from any subset of size $< n$ it is not possible to infer
anything about the positions of the remaining points.
\end{itemize}
For this measure,
the authors showed analytically that the upper limit of LM dimension   scales linearly with the number of parameters in a classical neural network with a factor of proportionality that is the unity.
In practice, a training method cannot be perfect, therefore this linear dependence persists with a lower factor of proportionality, 
For more details and the informational meaning of this measure we refer the interested reader to the original work of \cite{LM}. 

For quantum models where analytical calculations are not available, we proceed with
the numerical evaluation of LM dimension, which we denote as $\Tilde{D}_{LM}$. 
Naturally $\Tilde{D}_{LM}$ lower bounds $D_{LM}$ and this can be understood from  the procedure that we follow for each encoding-rotation model under test:
\begin{itemize}
\item We set the $k$-dimension of the inputs of the model. According to our general model for 
 a qudit, we have $k\leq d^2-1$. We generate a set of $\left\{ x_n  \right\}$   points in random position, which we call \textit{random pattern},  by selecting each of $k$ coordinates from a uniform distribution in the interval $\left[-0.5, 0.5\right]$.
 We start with a $n\leq P$ where $P=S+W$ the total number of parameters.
\item According to the definition of LM dimension, we treat only binary classification tasks, and we attribute labels randomly
to the vectors of the random pattern to two groups. For a given random pattern,  one should test all $2^n$ different labelings, but not having the computational capacity for this, when $n>6$ we perform our estimate by taking a sample of $50$ different random labelings.
\item If the training of parameters via gradient descent, with $50$ different starting points,  does not lead to   classification of the vectors of the random pattern into two groups  with $100\%$ success ratio, we repeat for other random patterns $\left\{ x_n  \right\}$   until
we find a pattern that is successfully classified for all possible labelings. However, we do not exceed the number of $10$ different random patterns under test.
\item The number $n$ is step-wised increased up to the point where the classification is no longer
 successful for any tested random pattern. The empirical LM dimension, $\Tilde{D}_{LM}$  is the highest $n$ where the classification is achieved for at least one random pattern (all possible labelings).
\end{itemize}

In the Table~\ref{Tab1} we present the results on the empirical estimation of LM dimension
for a qubit. The  generators of the algebra, $\hat{g}_i$ for a qubit system, are identified as the Pauli operators:
$ \hat{g}_1=\hat{\sigma}_x, ~\hat{g}_2=\hat{\sigma}_y,~\hat{g}_3=\hat{\sigma}_z ~~.$ For all qubit schemes under study the classification is performed by measuring the operator $\hat{g}_3$ with eigenvalues $\left\{-1,1\right\}$ and corresponding eigenvectors $\left\{\left|1\right\rangle,\left|0\right\rangle\right\}$. The two groups of data are separated according $\left\langle \hat{g}_3\right\rangle \gtrless 0$. For comparison with our \textit{single-layer} model, we have also included 
models  ($E-F$ Table~\ref{Tab1}) which implement  re-uploading  of input data \cite{Latore, Heid}. 

\begin{widetext}
\begin{table}[h]
\caption {Qubit models} \label{Tab1} 
\begin{center}
    \begin{tabular}{ |c| l | l | c | c |}
    \hline
     Model & Encoding-Rotation & k & P=S+W   & $\Tilde{D}_{LM}$ \\ \hline  \hline
		A & $\exp\left[i s_1 \left( x_1 \hat{g}_1 +i x_2 \hat{g}_2\right)\right] \exp\left[i w_1  \hat{g}_1 \right]$   & 2 & 2 & 4 \\ \hline
  B & $\exp\left[i s_1 \left( x_1 \hat{g}_1 +i x_2 \hat{g}_2\right)\right] \exp\left[i w_1  \hat{g}_1 +i w_2 \hat{g}_2+i w_3 \hat{g}_3\right]$   & 2 & 4 & 7 \\ \hline
  C & $\exp\left[i  \left( x_1 \hat{g}_1 +i x_2 \hat{g}_2\right)\right] \exp\left[i w_1  \hat{g}_1 +i w_2 \hat{g}_2+i w_3 \hat{g}_3\right]$   & 2 & 3 & 3 \\ \hline
   D & $\exp\left[i s_1 x_1 \hat{g}_1 +i s_2 x_2 \hat{g}_2\right] \exp\left[i w_1  \hat{g}_1 +i w_2 \hat{g}_2+i w_3 \hat{g}_3\right]$   & 2 &5 & 6 \\ \hline
  E & $\exp\left[i s_1 \left( x_1 \hat{g}_1 +i x_2 \hat{g}_2\right)\right] \exp\left[i w_1  \hat{g}_1 \right]$  &  & &  \\ \hline
     & $\exp\left[i s_2 \left( x_1 \hat{g}_2 +i x_2 \hat{g}_3\right)\right] \exp\left[i w_2  \hat{g}_2 \right]$  & 2 & 4 & 8 \\ \hline
      F & $\exp\left[i s_1 \left( x_1 \hat{g}_1 +i x_2 \hat{g}_2\right)\right] \exp\left[i w_1  \hat{g}_1 \right]$  &  & &  \\ \hline
     & $\exp\left[i s_2 \left( x_1 \hat{g}_2 +i x_2 \hat{g}_3\right)\right] \exp\left[i w_2  \hat{g}_2 \right]$  &  &  &  \\ \hline
      & $\exp\left[i s_3 \left( x_1 \hat{g}_1 +i x_2 \hat{g}_2\right)\right] \exp\left[i w_3  \hat{g}_1 \right]$  & 2 & 6 & 9 \\ \hline
    G & $\exp\left[i s_1 \left( x_1 \hat{g}_1 +i x_2 \hat{g}_2+i x_3 \hat{g}_3\right)\right] \exp\left[i w_1  \hat{g}_1 +i w_2 \hat{g}_2\right]$   & 3 & 3 & 6 \\ \hline 
    \end{tabular}    
\end{center}
\end{table}
\end{widetext} 

 We proceed with the estimation of $\Tilde{D}_{LM}$ for a qutrit with results presented in Table~\ref{Tab2}. The generators $\hat{g}_i$ of the $SU(3)$ group can be chosen to  be the  Gell-Mann operators, $\hat{\lambda}_i$ with $i=1,\ldots,8$,  which are provided in matrix form in the Appendix.
According to Section~\ref{method}, during the encoding phase the classical data are mapped onto the Bloch hyper-sphere of a qutrit embedded in the $8$-dimensional space which cannot be visualized.  To obtain a partial visualization, as for example in Section~\ref{3groups}, we use the Bloch-Ball representation offered by the  $su(2)$ subalgebra of $su(3)$, spanned by the generators $\left\{\hat{L}_x= \hat{\lambda}_1+\hat{\lambda}_6,~\hat{L}_y=\hat{\lambda}_2+\hat{\lambda}_7, \hat{L}_z=\hat{\lambda}_3+\sqrt{3}\hat{\lambda}_8\right\}$. 
For all schemes,  we choose to  measure the operator  $ \hat{L}_z=\hat{\lambda}_3+\sqrt{3}\hat{\lambda}_8$ that is
diagonal in the computational basis and with uniform spectrum $\left\{-2,0,2\right\}$. 
For binary-classification results, as shown in Table~\ref{Tab2}, we separate the two groups according to the sign of $\left\langle \hat{L}_z \right\rangle$.

\begin{widetext} 
\begin{table}[h]
\caption {Qutrit models} \label{Tab2} 
\begin{center}
    \begin{tabular}{ |c| l | l | c | c |}
    \hline
     Model & Encoding-Rotation & k & P=S+W   & $\Tilde{D}_{LM}$ \\ \hline  \hline
     A & $\exp\left[i  x_1 s_1 \hat{g}_6 +i x_2s_2 \hat{g}_7\right]$  $\exp\left[i  w_1 \hat{g}_1+i  w_2 \hat{g}_4\right] $  & 2 & 4 & 6 \\ \hline
		B  & $\exp\left[i  x_1( s_1 \hat{g}_1 +s_2 \hat{g}_2)+i  x_2( s_3 \hat{g}_3 +s_4 \hat{g}_4)\right]$  & &  &  \\ \hline
  & $\exp\left[i \sum_{j=1}^4 w_j \hat{g}_j\right] $  & 2 & 8 & 8 \\ \hline
  C & $\exp\left[i  x_1 ( s_1 \hat{g}_5 +s_2 \hat{g}_6)+i  x_2( s_3 \hat{g}_7 +s_4 \hat{g}_8)\right]$  & &  &  \\ \hline
   & $\exp\left[i \sum_{j=1}^4 w_j \hat{g}_j\right] $  & 2 & 8 & 7 \\ \hline
 D1 & $\exp\left[i   s_1\sum_{j=1}^8 x_j \hat{g}_j
\right]\exp\left[i \sum_{j=1}^7 w_j \hat{g}_j\right] $  & 8 & 8 & 13 \\ \hline
D2 & $\exp\left[i   \sum_{j=1}^8 x_j s_{1+jmod4} \hat{g}_j
\right]\exp\left[i \sum_{j=1}^4 w_j (\hat{g}_j+\hat{g}_{j+4})\right] $   & 8 & 8 & 16 \\ \hline
 D3 & $\exp\left[i   \sum_{j=1}^8  s_j x_j \hat{g}_j
\right]\exp\left[i  w_1 \hat{g}_1\right] $  & 8 & 9 & 17 \\ \hline
    \end{tabular}
\end{center}
\end{table}
\end{widetext}

With regard to efficiency, the single-layer schemes that achieve $\Tilde{D}_{LM}=2 P$ can be considered as the most successful ones, i.e. qubit: $A$, $G$, qutrit: $D2$.
 From the qubit models $C$,$D$ and qutrit model $D1$ we may conclude that both the absence and the excessive input of parameters in the encoding phase is not recommended. We also observe that the most successful single-layer
 models are the ones where $k \approx d^2-1$.
 
For classical neural networks, for a fixed input dimension $k$, one can linearly augment  LM dimension with the number of parameters by adding hidden layers \cite{LM}. For the model presented here,  this becomes possible by 
using a qudit system where $k<d^2-1$. The scaling $\Tilde{D}_{LM}=2 P$
is not maintained but one rather achieves $\Tilde{D}_{LM}\approx P$ as it is shown with $k=2$ with qutrit model $B$. We have also implemented classification with $k=2$ for a $4$-level system (not shown), where $k=2$, $P=13$ and $\Tilde{D}_{LM}=10$. An alternative way to increase LM dimension is to use re-uploading, see qubit models $E-F$, but there the scaling $\Tilde{D}_{LM}=2 P$ also  is not achieved but rather $\Tilde{D}_{LM}= P+L$, with $L$ being a constant. (We have also implemented $4$-layers re-uploading, extending models $E-F$ with $k=2$, $P=8$ and where we estimated  $\Tilde{D}_{LM}=11$ - not shown in the tables.) This analysis confirms the findings in \cite{QCapacity} and underlines the need for more research in identifying quantum models which exceed the classical limits.

Finally, for single-layer models and $k\approx d^2-1$, we see
that $\Tilde{D}_{LM}$ is higher than the one for the classical neural network.
It would be interesting to see whether more exotic classical perceptron models
such as product-units \cite{Durbin,dellenetal2019} or complex-valued perceptron \cite{BU} exhibit similar augmentation of LM dimension. 
In addition we underline the fact  that the LM dimension only captures a specific aspect of the model. For a complete evaluation of the quantum model for supervised learning task, 
other aspects \cite{QFisher} would have to be taken into account, e.g., difficulty in training (barren-plateaus problem), presence of noise in implementation etc.

The conclusions of the numerical studies on LM dimension are illustrated  with  examples in next sections.
Since LM dimension only concerns capacity of binary classification tasks, we
  address classifciation problems with $M>2$ classes as well.

\section{ Classification problems treated with a qubit}
We start the illustration of the suggested method by addressing two typical classification problems with a qubit ($d=2$).
Even though the power of a qubit has been extensively studied in the literature, this is the first
example showing that a qubit can be logically complete, i.e., it is able to  implement all binary logical functions. This is achieved with the model $A$, Table~\ref{Tab1}, which contains  two real parameters. 
 This outcome does not come as surprise since model $A$ has  LM dimension $\Tilde{D}_{LM}=4$ for $k=2$, or, in other words, can shatter all possible ways
 four $2$-dimensional vectors in random positions.

\subsection{Binary logical functions }
Let us consider four data-sets on a plane ($k=2$), as shown in Fig.1~(a).  
The logical functions for these noisy data correspond to different
attributions of each data-set to one of  two groups, $A$ and $B$.
For instance, the  XOR function requires a classification of the data-sets  as in  Fig.1~(a).

To implement classification according to the logical functions, we first map the data 
onto the 2-dimensional surface of the Bloch sphere. Even if the feature space has the same dimension as the initial space, the change in topology proves to be helpful.
Numerical tests show that all logical functions can be implemented this way with $2$ real weights ($S=1$ and $W=1$).
In more details, we use the encoding and rotation as in model-A for a qubit, see Table~\ref{Tab1}, and the classification is conducted using the sign of $\left\langle \hat{g}_3 \right\rangle$.
	
We successfully solved classification problems for all logical functions (AND, OR, XOR), however we present in Fig.1~(b) only the results about XOR, which is the most challenging task, since it is a non-linearly separable problem. 
The total number of data is $2000$ and we use $4\%$ of them for the training. A success ratio of classification of $100\%$ was readily achieved.

\begin{figure}[h]
\begin{center}
\includegraphics[width=5cm]{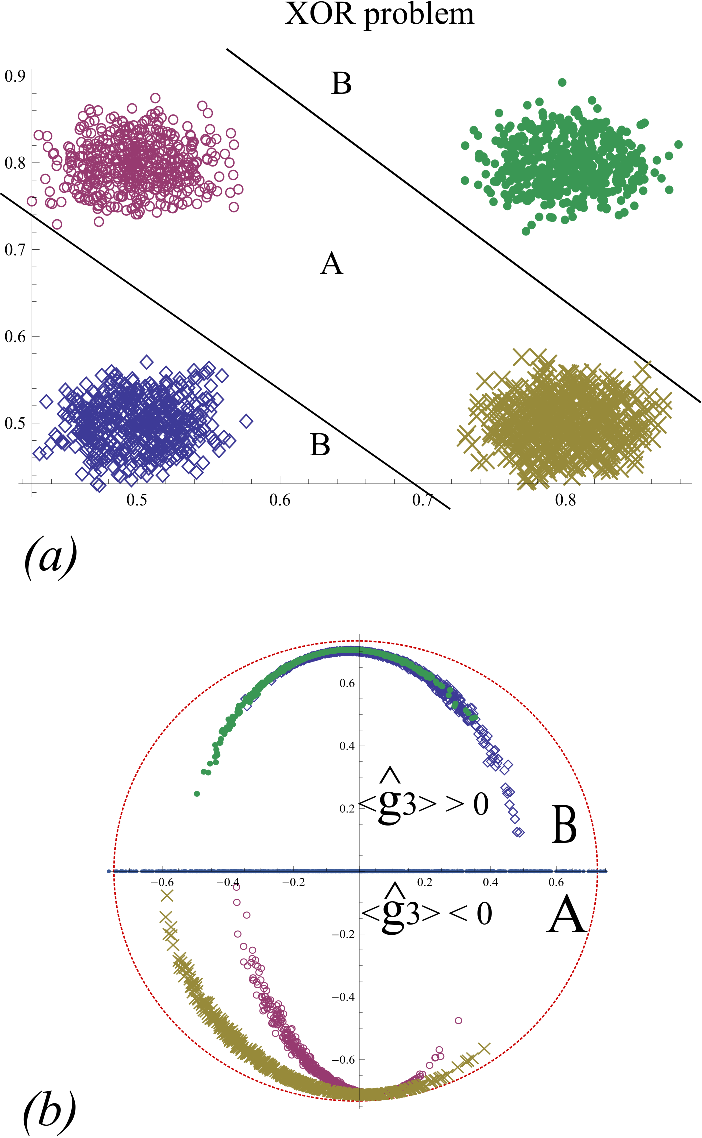}
\end{center}
\caption{\textit{(a)} Data to be classified according to XOR logical function, into  groups A and B. \textit{(b)} The classified data
mapped on the surface of Bloch sphere (projection on the x-z plane) after training on the $2$ weights has been performed.}%
\label{fig1}%
\end{figure}

It is important to note here that all binary logical functions
can be solved with $2$ real parameters also by the complex perceptron model presented in \cite{BU}. We proceed with an example that it is not solvable with any single-layer classical perceptron model up to our knowledge.
 


\subsection{Classification for circular boundaries}
We proceed with a more complex classification problem and show that it can still be tackled with a single qubit.
For this purpose, we employ model $B$, Table~\ref{Tab1}, because it achieves a higher LM dimension than model $A$.

\begin{figure}[h]
\begin{center}
\includegraphics[width=5cm]{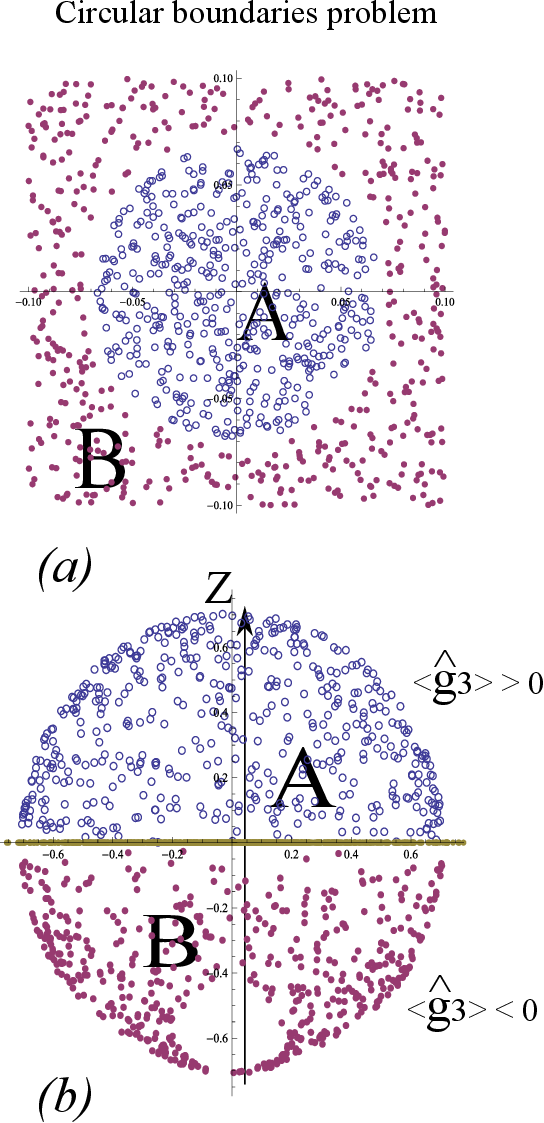}
\end{center}
\caption{\textit{(a)} The initial data (1000 points) to be classified into  groups A and B. \textit{(b)} The data are mapped on the Bloch surface and perfectly classified.}%
\label{fig3}%
\end{figure}

The problem  consists of classifying the data ($1000$ 2-dimensional vectors) in Fig.~\ref{fig3}~(a) into two groups. In Fig.~\ref{fig3}~(b), we present the classification achieved on the Bloch sphere after 
the weights $s_1, w_1, w_2, w_3$ have been optimized. The classification ratio achieved is $100\%$ using $10\%$ of total data set as training data set.

Following the same encoding-rotation scenario (model $B$), we are able to treat elliptical data (not presented here), but with a lower final classification ratio ($\approx 90\%$).

\section{ Examples solved with a qutrit}
Even though we have been able to solve  a couple of basic classification problems with one qubit, it is obvious that one needs a higher dimensional space $d$ to resolve more complicated problems since one qubit can
accommodate at most $5$ parameters/weights according to our single-layer model.  
As shown in the Section~\ref{LM}
qutrits may accomodate more parameters and therefore achieve higher LM dimension. In addition, tests have shown that qutrit models perform better than qubit models for classification tasks into $M>2$ groups. This is not obvious studying LM dimension alone. 

\subsection{Noisy XOR}
\begin{figure}[h]
\begin{center}
\includegraphics[
width=8cm]{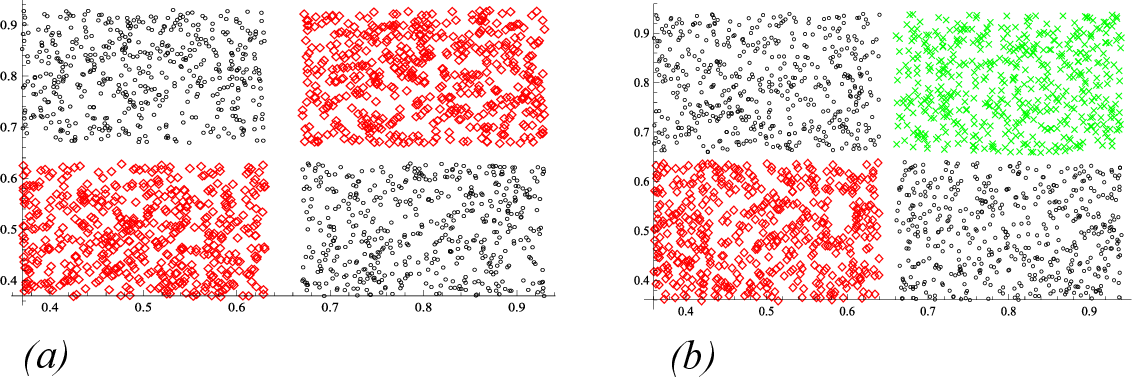}
\end{center}
\caption{Qutrit model: \textit{(a)}  classification into $2$ groups employing $8$ weights, \textsl{(b)}  classification into $3$ groups employing $9$ weights.}%
\label{fig4}%
\end{figure}

We first investigate  the binary classification  task presented in Fig.\ref{fig4}~(a) for which all qubit-models exhibited low performance but where qutrit's model $B$, see Table~\ref{Tab2}, gives adequate results. More specifically, 
 we use
$1\%$ of the total data ($2000$ points) for training  and achieve  a success  classification ratio of $96\%$.

\subsubsection{Classification into three groups and a geometric picture \label{3groups}}
We increase the difficulty of the previous problem by demanding classification in $3$ groups of data and reducing the margins between sets, as shown in Fig.4~(b).
We use a comparable number of weights ($9$), but now the encoding-rotation model is:
\begin{itemize}
	\item Encoding via
	\begin{eqnarray}
\left|\psi_{\vec{x}, \vec{s} }\right\rangle=& \exp\left[i  x_1 ( s_1 \hat{g}_3 +s_2 \hat{g}_5+s_3 \hat{g}_7)\right.\nonumber\\&\left.+i  x_2( s_4 \hat{g}_4 +s_5 \hat{g}_6+s_6 \hat{g}_8)\right]
 \left|0\right\rangle~~.
\end{eqnarray}
\item Rotation via
\begin{equation}
\left|\psi_{\vec{x}_l, \vec{s},\vec{w} }\right\rangle= \exp\left[i \sum_{j=1}^3 w_j \hat{L}_j\right]\left|\psi_{\vec{x}_l, \vec{s} }\right\rangle
\end{equation}
where $\hat{L}_1=\hat{L}_x,~\hat{L}_2=\hat{L}_y,~\hat{L}_3=\hat{L}_z$.
\item Measurement of 	$\hat{L}_z$ and classification
 by comparing  the value  of $\left\langle \hat{L}_z\right\rangle$ with $A:\left[-2, -2+4/3\right]$, $B:\left[ -2+4/3,2-4/3\right]$, $C:\left[ 2-4/3,2\right]$.
\end{itemize}.

\begin{figure}[h]
\begin{center}
\includegraphics[
width=5cm]{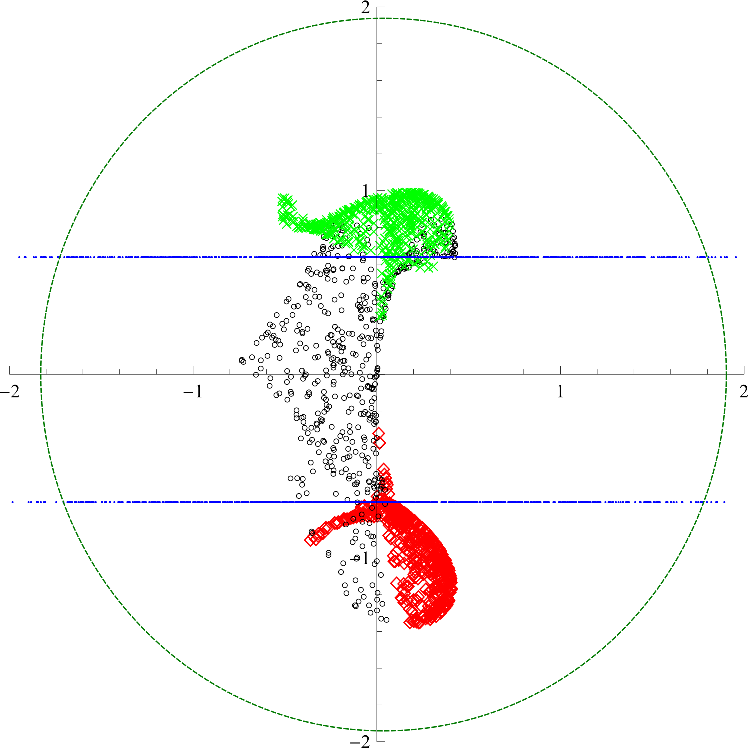}
\end{center}
\caption{The classification of data of Fig.~4 (b) into $3$ groups  as perceived in the $SU(2)$ Bloch sphere representation provided by the operators
$\left\{\hat{L}_x,\hat{L}_y\hat{L}_z\right\}$.}%
\label{fig5}%
\end{figure}

Using $4\%$  of  the total data ($2000$ points) for training, a success ratio of classification  $87\%$ is achieved for the rest of data.
In Fig.~\ref{fig5}, we depict the mapping (with optimized parameters) of the data into the $SU(2)$ Bloch ball generated by $\left\{\hat{L}_x,\hat{L}_y\hat{L}_z\right\}$ operators.
The classification `intervals' for $\left\langle \hat{L}_z\right\rangle$ are also presented in the picture as horizontal lines.
This `local' picture offered by the subgroup is equivalent to the picture one would obtain  by inspecting the local density matrix 
of an entangled system. One can thus claim by borrowing terms by the notion of generalized entanglement \cite{Viola} 
 that the self-entanglement of a qutrit has the same use in the classification procedure as physical entanglement between subsystems, i.e., this  extends the mapping from the surface to the inside area of the Bloch hypersphere of a subsystem. The generation of self-entanglement in a qudit does require the ability to fully operate the system but in practice this is less demanding than the entangling interaction between subsystems.

\subsection{Classifying moon sets with a qutrit}
\begin{figure}[h]
\begin{center}
\includegraphics[
width=6cm]{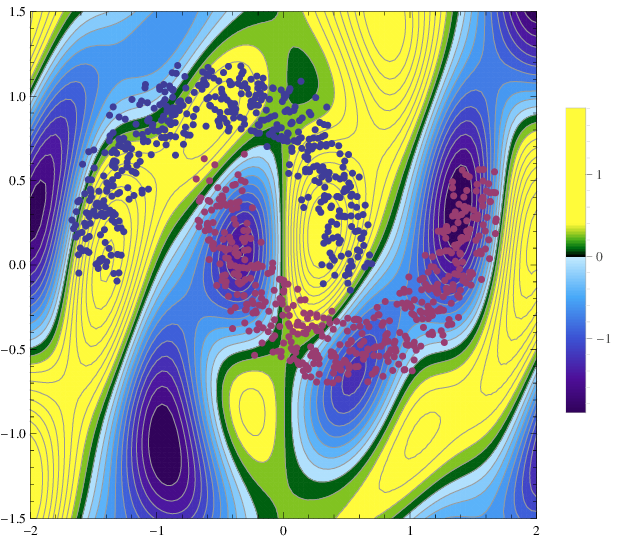}
\end{center}
\caption{Classification of moon sets with the qutrit model $C$ using $8$ weights. A contour plot of $\left\langle \hat{L}_z\right\rangle$ is depicted together with the moon data after optimization has been performed.}%
\label{fig6}%
\end{figure}

Finally, by using qutrit model $C$ of Table~\ref{Tab2}, we attempt a common  classification task, the one of moon sets.
By optimizing the $8$ parameters of the model, we achieve a classification ratio of $90\%$ using  $10\%$ of $800$ total data points.
In Fig.~\ref{fig6}, we present $\left\langle \hat{L}_z\right\rangle$ for the optimized set of parameters, together with the data sets.

\subsection{Real-world data multi-class classification \label{rwd}}
We will now turn to multi-class classification tasks using real-world data
and more advanced methods of training. We use data sets from the UCI Machine Learning Repository, a widely-used and publicly available repository \cite{uci}, maintained by the University of California, Irvine.  Our aim is to explore the feasibility of using a single qutrit to accurately distinguish between three classes in data sets with more than two dimensions, such as the Iris and Wine datasets. Our results illustrate that supervised learning in the  context of less structured data is achievable.

The Iris dataset consists of 150 samples of iris flowers, with measurements of four features: sepal length, sepal width, petal length, and petal width. Each sample is labeled with one of three possible iris species. The Wine Cultivars dataset consists of measurements of thirteen chemical constituents found in three different wine cultivars. The objective is to classify the cultivar of the wine based on the chemical composition measurements. 

\begin{table}[h]
\caption {Real-world data treated with a qutrit} \label{Tab3} 
\begin{center}
\begin{tabular}{|l|l|l|l|l|}
\hline
Dataset            & Features & Classes & Samples & Pre-processing \\ \hline
IRIS    & 4        & 3       & 150     & none          \\ \hline
WINE    & 13       & 3       & 178     & PCA           \\ \hline
\end{tabular}
\end{center}
\end{table}

Our aim is to use the same encoding and number of parameters for both data sets. Thus for the Wine Cultivars data set which possesses  $13$ different features we employ Principal Component Analysis (PCA) \cite{PCA} in order to reduce the number of features to four.  

The encoding and rotating scheme that we follow is 
\begin{itemize}
	\item Encoding via
	\begin{equation}
\left|\psi_{\vec{x}_l, \vec{s} }\right\rangle= \exp\left[i s \sum_{j=1}^4 x_j  ~ \hat{g}_{j} \right]
 \left|0\right\rangle~~.
\end{equation}
\item Rotation via
\begin{equation}
\left|\psi_{\vec{x}_l, \vec{s},\vec{w} }\right\rangle= \exp\left[i \sum_{j=5}^8 ~ w_{j-4}  ~ \hat{g}_8\right]\left|\psi_{\vec{x}_l, \vec{s} }\right\rangle
\end{equation}
where the variational weights $\vec{w} = (s, w_1, w_2, w_3,w_4)$ are the parameters to be optimized.
\end{itemize}

We ensured an equal representation of each class. For re-producibility, we used the same seed to split the data into train and test sets. To avoid overfitting, early stopping was employed and different gradient-based methods were trialed to combat the barren plateaus problem before settling to stochastic gradient descent (SGD) \cite{gradient} using the parameter-shift rule.  This method reduces the number of measurements needed during implementation compared to the standard method, making it more efficient and practical for quantum machine learning. SGD is a variant of gradient descent that randomly selects a subset of data points, called mini-batch, to calculate the gradient of the cost function at each iteration. Since these are multi-class problems categorical cross entropy loss was used as a cost function. Using this approach, we achieved competitive scores with a single qutrit as can be seen in the Table~\ref{Tab4}.

Since these are multi-class problems categorical cross entropy was used as cost function, which combines the softmax activation and the negative log likelihood loss as follows:
\begin{equation}
\text{{loss}} = -\frac{1}{N} \sum_{i=1}^{N} \sum_{j=1}^{C} t_{ij} \log(p_{ij})~.
\end{equation}
Here $N$ is the number of samples, $C$ is the number of classes, $t_{ij}$ represents the true label for sample $i$ and class $j$, and $p_{ij}$ represents the predicted probability.

Using this approach, we achieved competitive scores with a single qutrit as can be seen in the Table~\ref{Tab4}.

\begin{table}[h]
\caption {Comparative numerical studies for  classification of Iris and Wine data. The train set accuracy (TrSA) and test set accuracy (TeSA) reached with different methods. } \label{Tab4} 
\begin{center}
\begin{tabular}{ | l | l | l | l |}
    \hline
                        & Classical Model & Entangled Qubits & Single Qutrit \\
                        \hline
Iris TrSA & 97.5\%          & 85\%             & 86.67\%        \\
Iris TeSA  & 100\%           & 86.6\%           & 84.44\%          \\
Wine TrSA & 97.9\%          & 65.5\%           & 77.42\%        \\
Wine TeSA  & 91.7\%          & 69.4\%           & 85.19\%   
\\
\hline
\end{tabular}
\end{center}
\end{table}

In these benchmarks, we present the results of the single qutrit model against a classical machine learning model using Support Vector Machines (SVM) and a Variational Quantum Classifier (VQC) model with entangled qubits in Qiskit. These tests were conducted using four qubits and the popular ZZ feature map with twelve parameters, utilizing the Limited-memory Broyden-Fletcher-Goldfarb-Shanno Bound (L-BFGS-B) optimizer to minimize sensitivity to local minima and the barren plateau issue \cite{lbfgsb}.

These results showcase that even a single-qudit classifier is capable of multi-class classification for multi-dimensional real-world data. Although on the Iris data set the four-qubits model outperformed the single qutrit model, the single qutrit model produced better results even with five parameters compared to the twelve used by the ZZ feature map on the Wine data set. Increasing the encoding layers could further enhance the classifier's performance, but since the aim of our study was to demonstrate that a single-layer qudit classifier can accurately distinguish between multiple classes, it is not further investigated here.

\section{Discussion}
Qudits are extensions of qubit units to higher-dimensions,
which can enhance the performance in quantum computing \cite{qu3, Floratos, speedup} and communication \cite{QC, QKD, arnault}. These are experimentally realizable with different physical models and recent proposals also use them in quantum machine learning \cite{Heid, qudit}.
In this work, we have described a model for data classification using a single  qudit. The parametrization is introduced according
to geometric intuition, partially for controlling the mapping on the Bloch hyper-surface and partially for adjusting 
the projective measurement to the data set's orientation on the Bloch hyper-sphere. Entangling or  adding more layers can certainly enhance the quantum classifier, similar to how classical neural networks yield better results with increased depth. Nonetheless, given the expense and error-prone nature of entangling in near-term quantum hardware, our results indicate that even a low-depth single-qudit classifier holds a promise for quantum machine learning,  if it is thoughtfully employed with a balanced distribution of parameters in the encoding and rotating steps.

The simple model that we present shares obvious similarities and borrows ideas from previous works \cite{Latore, Farhi, Harrow, Heid}.
Being though only in the mid-way of exploration of the potential role of  quantum systems for ML tasks, this geometrically-dressed entanglement-free proposal gives its own contribution, connecting current efforts with the geometry of Hilbert-Schmidt space and underlying the equivalence of self-entanglement \cite{Viola} with physical one in practice. In addition, with the help of empirical estimation of LM dimension for a qubit and a qutrit,
 we have been able to demonstrate that the `capacity' of single-layer quantum systems can be higher 
 than for classical neural network systems bearing the same number of training parameters. It remains an open question for future work to investigate and compare the capacity of the quantum model with more intrigued single-layer classical perceptron models but also
 to investigate whether quantum multi-layer structures can exist  which can keep the advantage in LM dimension over classical NN.

\section*{Acknowledgements}
 AM and DS acknowledge partial support  by the European Union’s Horizon Europe research and innovation program under grant agreement No.101092766 (ALLEGRO Project).

\appendix 
\begin{widetext}

\section{Qubit Kernel}

There is the common belief  that the encoding of classical data via rotation angles results in 
a simple cosine kernel, that is easily classically reproducible. However cosine kernels only emerge  when the rotation encoding
is successive as
\begin{equation}
\exp\left[i s x_1 \hat{g}_1 \right]\exp\left[i s x_2 \hat{g}_2\right]\ldots \left|0\right\rangle~~,
\end{equation}
or concerns a setting of non-interacting qubits,
\begin{equation}
\exp\left[i s x_1 \hat{g}_1 \right] \left|0\right\rangle \otimes \exp\left[i s x_2 \hat{g}_2\right] \left|0\right\rangle\ldots ~~.
\end{equation}

In this work we use rotation encoding that looks very similar
\begin{equation}
\left| \psi_{\vec{x}, s }\right\rangle=\exp\left[i s ( x_1 \hat{g}_1 +x_2 \hat{g}_2)\right]\left|0\right\rangle \label{quk}
\end{equation}
staying linear in the input $\vec{x}$
but whose corresponding kernel is more intrigued than cosine one due to BCH formula.
Straightforward calculations show that for qubit model $A$  the kernel writes as
\begin{eqnarray}
 \left|\left\langle \psi_{\vec{x}, s} \right.\left|\psi_{\vec{y},s }\right\rangle\right|^2 &=& \frac{1}{8 x^2
   y^2}\left(2 x^2 y^2 \cos (2 s y)+x^2 y^2 \cos
   (2 s (x+y)) \right. \nonumber\\
	& + &\left(x^2
   \left(y^2-y_2^2\right)-x_2^2 y^2+2 x_1
   x y y_1+\left(x_1 y_1+x_2
   y_2\right){}^2\right) \cos (2 s  (x-y))\nonumber\\
	&+&2 \left(x^2
   \left(y^2+y_2^2\right)-x_2^2
   y^2-\left(x_1 y_1+x_2
   y_2\right){}^2\right) \cos (2 s x)-2
   x^2 y_2^2 \cos (2 s y)-x^2 y_2^2 \cos
   (2 s (x+y))\nonumber\\
   &+&2 x_2^2 y^2 \cos (2 s
   y)-x_2^2 y^2 \cos (2 s (x+y))
	-   2 x x_1
   y_1 y \cos (2 s (x+y))-2 x_1^2 y_1^2
   \cos (2 s y) \nonumber\\
   &+&x_1^2 y_1^2 \cos (2 s (x+y))-2 x_2^2 y_2^2 \cos (2 s
   y)+x_2^2 y_2^2 \cos (2 s (x+y))-  4 x_1
   x_2 y_1 y_2 \cos (2 s y) \nonumber\\ 
    &+& \left. 2 x_1 x_2 y_1
   y_2 \cos (2 s (x+y))+2 x^2 y^2+2 x^2 y_2^2+2 x_2^2 y^2+2 x_1^2 y_1^2+2
   x_2^2 y_2^2+4 x_1 x_2 y_1 y_2\right)
\end{eqnarray}
where $x=\sqrt{x_1^2+x_2^2}$ and $y=\sqrt{y_1^2+y_2^2}$.
Naturally, the intricacy of kernels emerging in this work  is increasing with the dimension of the input $k$ and dimension $d$.

\section{ Gell-Mann operators}
Here we list  the generators of $su(3)$ algebra, the so called  Gell-Mann operators, as matrices in the computational
basis of a qutrit, i.e., $\left\{\left|0\right\rangle,~\left|1\right\rangle,~\left|2\right\rangle\right\}$,
\begin{eqnarray*}
\hat{\lambda}_1=\left(
\begin{array}{ccc}
 0 & 1 & 0 \\
 1 & 0 & 0 \\
 0 & 0 & 0 \\
\end{array}
\right),~ & \hat{\lambda}_2=\left(
\begin{array}{ccc}
 0 & -i & 0 \\
 i & 0 & 0 \\
 0 & 0 & 0 \\
\end{array}
\right),~  &\hat{\lambda}_3=\left(
\begin{array}{ccc}
 1 & 0 & 0 \\
 0 & -1 & 0 \\
 0 & 0 & 0 \\
\end{array}
\right),~ \\ \hat{\lambda}_4=\left(
\begin{array}{ccc}
 0 & 0 & 1 \\
 0 & 0 & 0 \\
 1 & 0 & 0 \\
\end{array}
\right),~ & \hat{\lambda}_5=\left(
\begin{array}{ccc}
 0 & 0 & -i \\
 0 & 0 & 0 \\
 i & 0 & 0 \\
\end{array}
\right),~ & \hat{\lambda}_6= \left(
\begin{array}{ccc}
 0 & 0 & 0 \\
 0 & 0 & 1 \\
 0 & 1 & 0 \\
\end{array}
\right),~ \\ \hat{\lambda}_7=\left(
\begin{array}{ccc}
 0 & 0 & 0 \\
 0 & 0 & 1 \\
 0 & 1 & 0 \\
\end{array}
\right),~ & \hat{\lambda}_8=\frac{1}{\sqrt{3}}\left(
\begin{array}{ccc}
 1 & 0 & 0 \\
 0 & 1 & 0 \\
 0 & 0 & -2 \\
\end{array}
\right).~  & 
\end{eqnarray*}
For the extended set one should add $\hat{g}_0=\sqrt{\frac{2}{3}}\hat{1}$.
\end{widetext}

\end{document}